\documentclass{article}
\usepackage[cp1251]{inputenc}
\usepackage[dvips]{graphicx}
\usepackage{amssymb}
\begin{document}
\title{A possibility of increasing spin injection efficiency\\ in magnetic junctions}
\author{E.\,M.\,Epshtein, Yu.\,V.\,Gulyaev,
P.\,E.\,Zilberman\/\thanks{Corresponding author. E-mail: zil@ms.ire.rssi.ru}, A. I. Krikunov \\ \\
        Institute of Radio Engineering and Electronics\\ of the Russian Academy of
        Sciences,\\ Fryazino, 141190 Russia}

\date{}
\maketitle

\begin{abstract}
Nonequilibrium electron spin polarization is calculated under spin
injection from one ferromagnet to another in magnetic junction. It is
shown that the nonequilibrium spin polarization can be comparable with
equilibrium one if the material parameters are chosen appropriately. This
leads to lowering the threshold current density necessary for the junction
switching and opens a perspective to creating THz laser based on the
spin-polarized current injection.
\end{abstract}

%PACS 72.25.Ba, 72.25.Hg, 75.47.-m
The pioneer work by Aronov and Pikus~\cite{Aronov} marked the beginning of
vast literature devoted to the problem of spin injection under current
flowing through the interface between a ferromagnet ($F$) and nonmagnetic
conductor ($N$) or another ferromagnet~\cite{Son}--\cite{Qi}. Such an
attention is due to several reasons. First, there is a perspective of
creating spin analogs of transistor type semiconductor devices in which
charge injection is replaced by spin injection. Spin injection determines
also the giant magnetoresistance effect in spin-valve type systems (see,
e.g.,~\cite{Schmidt2,Fert,Yu2,Dieny}) and the fluctuation instability
effect in such systems under high enough current density through the
contact between ferromagnets that leads to switching from antiparallel
configuration to parallel one~\cite{Gulyaev1,Gulyaev2}. Finally, if the
spin subband population inversion by injection is reached, the amplifiers
and generators for $10^{12}$--$10^{13}$ Hz frequency range may be created.
In this connection, note Ref.~\cite{Osipov}, where spin injection effect
on the microwave absorption and emission in nonmagnetic semiconductor
n-InSb was observed, as well as Ref.~\cite{Kadigrobov}, in which a
possibility is discussed of creating a solid-state THz laser based on the
spin-polarized electron tunnel injection from one ferromagnet to another.

In the present paper we try to show that a significant (by several orders
of magnitude) increasing the injected nonequilibrium spin density is
possible under appropriate parameter choice. Such a possibility was not
perceived in Refs.~\cite{Son}--\cite{Qi}, because symmetric $F/N/F$ type
structures were considered there with collinear relative orientation of
the same ferromagnetic layers.

Let us consider a $F_1/F_2/N$ system consisting of three contacting
layers~--- a semi-infinite ferromagnetic layer 1 in $-\infty<x<0$ range, a
ferromagnetic layer 2 of finite thickness $L$ in $0<x<L$ range, and a
semi-infinite nonmagnetic layer 3 in $L<x<\infty$ range. When electric
current flows in the $x$ axis positive direction ($1\rightarrow
2\rightarrow 3$), the spin polarization degree changes near the
interfaces. The measure of the spin polarization is
$P(x)=(n_+(x)-n_-(x))/n$, where $n_\pm(x)$ are partial spin-up and
spin-down electron densities, $n=n_+(x)+n_-(x)$ is total electron density
that is assumed to be constant (independent of $x$) because of neutrality
condition. The polarization deviation $\Delta P(x)=P(x)-\bar P$ from the
equilibrium value $\bar P$ in each of the layers is described by a
steady-state spin-current continuity equation~\cite{Gulyaev3}
\begin{equation}\label{1}
  \frac{dJ(x)}{dx}+\frac{\hbar n}{2\tau}\Delta P(x)=0,
\end{equation}
where $\tau$ is effective longitudinal spin relaxation time, which is
related with corresponding partial quantities by
$\tau^{-1}=\tau_+^{-1}+\tau_-^{-1}$, $J(x)$ is spin current density
defined as $J(x)=(\hbar/2e)\left(j_+(x)-j_-(x)\right)$, where $j_\pm(x)$
are partial electric current densities, $j=j_+(x)+j_-(x)$ is total
electric current density that is assumed to be constant because of the
one-dimensional geometry of the problem.

In the two spin subband model, the partial currents take the form
\begin{equation}\label{2}
  j_\pm=e\mu_\pm n_\pm(x)E(x)-eD_\pm\frac{dn_\pm(x)}{dx},
\end{equation}
where  $E(x)$ is local electric field, $\mu_\pm$ and $D_\pm$ are partial
electron mobilities and diffusion constants, respectively; they are
assumed to be invariable under breaking spin equilibrium.

By expressing $E(x)$ in terms of the current density $j$ and taking the
$n_\pm(x)=(n/2)\left(1\pm P(x)\right)$ relation into account, we obtain
the following formula for the spin current:
\begin{eqnarray}\label{3}
  &&J(x)=(\hbar/2e)\left[1+(en/2\sigma)(\mu_+-\mu_-)\Delta
  P(x)\right]^{-1}\nonumber\\
  &&\times\biggl\{\left[Q+(en/2\sigma)(\mu_++\mu_-)\Delta
  P(x)\right]j\nonumber\\
  &&-en\left[\tilde D+(en/2\sigma)(\mu_+D_--\mu_-D_+)\Delta
  P(x)\right]\frac{d\Delta P(x)}{dx}\biggr\},
\end{eqnarray}
where $\sigma=\sigma_++\sigma_-$ and $\sigma_\pm=e\mu_\pm\bar n_\pm$ are
the total and partial conductivities in spin equilibrium state,
respectively, $\bar n_\pm$ are equilibrium partial electron densities in
the spin subbands, $Q=(\sigma_+-\sigma_-)/\sigma$ is the conduction spin
polarization, $\tilde D=(D_+\sigma_-+D_-\sigma_+)/\sigma$ is effective
spin diffusion constant.

At $|\Delta P|\ll 1$ we have in linear approximation
\begin{equation}\label{4}
  J(x)=\frac{\hbar}{2e}\left\{Qj-en\tilde D\frac{d\Delta
  P(x)}{dx}+\frac{en\tilde\mu}{\sigma}j\Delta P(x)\right\},
\end{equation}
where $\tilde\mu=(\mu_+\sigma_-+\mu_-\sigma_+)/\sigma$ is effective spin
mobility. It can be shown easily that the ratio of the third summand in
Eq.~(\ref {4}) to the second one has an order of  $j/j_D$, where
$j_D=enl/\tau$ is diffusion current density. At the parameter values
typical for metals we have $j_D\sim 10^{10}\,\rm{A/cm^2}$, so that
$j/j_D\ll 1$ at attainable current densities, and the last (drift) term in
Eq.~(\ref {4}) can be neglected. Then the substitution of Eq.~(\ref {4})
into~(\ref{1}) gives the steady-state diffusion equation for
nonequilibrium spins:
\begin{equation}\label{5}
  \frac{d^2\Delta P(x)}{dx^2}-\frac{\Delta P(x)}{l^2}=0,
\end{equation}
where $l=\sqrt{\tilde D\tau}$ is spin diffusion length.

Equations~(\ref{4}) and (\ref{5}) have been derived in the approximation
of small deviation from spin equilibrium ($|\Delta P|\ll
1$)~\cite{Gulyaev3}. Note that Eqs.~(\ref{4}) (without the drift term) and
(\ref{5}) retain their form at any deviation from spin equilibrium also
under some often used model assumptions (the same partial mobilities and
diffusion constants, neglecting the effect of spin equilibrium breakdown
on the partial conductivities).

The following boundary conditions take place under spin-polarized current
flowing through interface between two ferromagnets 1 and 2 in
$1\rightarrow 2$ direction~\cite{Gulyaev3,Gulyaev2}:
\begin{equation}\label{6}
  J_1\cos\chi=J_2,
\end{equation}
\begin{equation}\label{7}
  N_1\Delta P_1=N_2\Delta P_2\cos\chi,
\end{equation}
where $N=(n/2)\left(g_+^{-1}+g_-^{-1}\right)$, $g_\pm$ are partial
densities of states at the Fermi level in spin subbands, $\chi$ is the
angle between the magnetization vectors of the contacting layers.

Presence of the $\cos\chi$ multiplier in the boundary conditions is due to
change of the quantization axis under electron passing from one magnetic
layer to another. Under passing from a magnetic layer to nonmagnetic one
(or vise versa), there is no such change, so that $\cos\chi=1$  is to be
put.

The solution of Eq.~(\ref{5}) with boundary
conditions~(\ref{6}),~(\ref{7}) describing the distribution of the
nonequilibrium spin polarization $\Delta P(x)$ in layer 2 takes the form
~\cite{Gulyaev2}
\begin{eqnarray}\label{8}
  &&\Delta
  P(x)=(j/j_{D2})\biggl[\sinh\lambda+\nu_{23}\cosh\lambda
  +(1/\nu_{12})\cos^2\chi(\cosh\lambda+\nu_{23}\sinh\lambda)\biggr]^{-1}\nonumber\\
  &&\times\biggl\{(Q_1\cos\chi-Q_2)[\cosh(\lambda-\xi)+\nu_{23}\sinh(\lambda-\xi)]\nonumber\\
  &&+Q_2\left(\cosh\xi+(1/\nu_{12})\cos^2\chi\sinh\xi\right)\biggr\},
\end{eqnarray}
where $\lambda=L/l_2$, $\xi=x/l_2$. The parameters
$\nu_{12}=(j_{D2}/j_{D1})(N_1/N_2)$ and
$\nu_{23}=(j_{D3}/j_{D2})(N_2/N_3)$ describe the spin current matching in
the interface. They may be presented as a ratio of "spin
resistances"~\cite{Gulyaev4}:
\begin{equation}\label{9}
  \nu_{ik}=\frac{Z_i}{Z_k},\qquad Z=\frac{\rho l}{1-Q^2},\qquad\rho=\frac{1}{\sigma}.
\end{equation}

At $Z_1\gg Z_2$ the cathode layer 1 works as an ideal injector, in which
the spin polarization is equilibrium ($\Delta P=0$), while the spin
equilibrium breakdown occurs in the anode layer 2. In the opposite case,
$Z_1\ll Z_2$, an ideal collector regime takes place, when the spin
equilibrium is disturbed in layer 1 and remains unchanged in layer 2.

In Refs.~\cite{Son}--\cite{Qi} the injection from ferromagnetic metal to
semiconductor was considered, when $Z_1\ll Z_2$, as well injection from a
ferromagnetic metal to the same metal with antiparallel orientation of the
magnetic moment ($Z_1=Z_2$, $\chi=180^\circ$). In those cases, $\Delta
P\le j/j_D\ll 1$, so that the injection efficiency is rather
low~\cite{Schmidt1}.

Let us consider the case of thin layer 2 ($\lambda\ll 1$). The
Eq.~(\ref{8}) takes the form
\begin{equation}\label{10}
    \Delta P(x)=\frac{j}{j_{D2}}\frac{Q_1\nu_{12}\cos\chi}{\nu_{12}\nu_{23}+\cos^2\chi}.
\end{equation}

\begin{figure}
\includegraphics{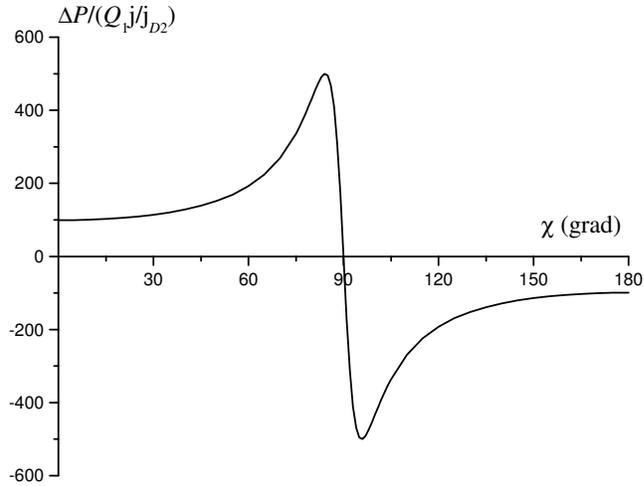}
\caption{The dependence of the nonequilibrium electron spin polarization
on the angle $\chi$ at $\nu_{12}=100$, $\nu_{23}=0.0001$.}\label{fig}
\end{figure}

As it was noted, $j/j_{D2}$ ratio is small usually. However, the second
multiplier in the right-hand side of Eq.~ (\ref{10}) can take large value
at $\nu_{12}\gg 1$, $\nu_{12}\nu_{23}\ll 1$ (or $Z_2\ll Z_1\ll Z_3$).
Therefore, in spite of the smallness of the $j/j_{D2}$  ratio, the $\Delta
P$ quantity can be comparable with $\bar P$. Such an effect allows a
simple physical interpretation. Under fulfillment of the conditions
indicated, effective injection from layer 1 to layer 2 takes place, while
the injection from layer 2 to layer 3 is ``locked''. The dependence of
$\Delta P$ on $\chi$ has the form of a resonance line in that case (see
Fig.~\ref{fig}). At $j\sim 10^7\,\rm{A/cm^2}$, $j_{D2}\sim
10^{10}\,\rm{A/cm^2}$, $\nu_{12}=100$, $\nu_{23}=0.0001$ we have $|\Delta
P|\sim 10^{-1}$ that is comparable with the equilibrium electron spin
polarization in ferromagnets~\cite{Moodera}.

To realize the conditions $Z_2\ll Z_1\ll Z_3$ corresponding to high spin
injection efficiency, a half-metal~\cite{Haghiri} may be chosen for the
layer 1, in that case spin resistance $Z_1$ is high because the conduction
spin polarization $Q_1$ is close to 1 (see Eq.~(\ref{9})). Semiconductor
with large $Z_3$ value may be taken for the nonmagnetic layer 3. Large
$Z_3$ value may be achieved in the case due to high resistivity and larger
spin diffusion length. The layer 2 in which nonequilibrium spin
polarization appears may be a ferromagnet such as cobalt.

An important consequence of the increasing spin injection efficiency is
lowering the threshold of the lattice magnetization fluctuation
instability under spin-polarized current flowing through the magnetic
junction. Indeed, the \emph{sd} exchange interaction effective field is
proportional to a derivative of the integral of $\Delta P(x)$ over the
layer thickness with respect to $\cos\chi$~\cite{Gulyaev1,Gulyaev2}.
Therefore, increasing of $\Delta P$  at $\chi=0$ and $\chi=180^\circ$
under dominating injection instability mechanism leads to lowering the
instability threshold and, consequently, the current density necessary for
switching the junction antiparallel configuration to parallel
one~\cite{Gulyaev2}. Besides, as is seen from the figure, $\Delta P(x)$
changes quickly near $\chi=90^\circ$ under sweeping the angle $\chi$.
Therefore, it may be expected that the switching will be much quicker
under the conditions indicated.

The work was supported by Russian Foundation for Basic Research, Grant No.
06-02-16197.

\end{document}